\documentclass[letter,twocolumn, nofootinbib,showkeys,showpacs,10pt]{revtex4}
\usepackage{amssymb}
\usepackage{amsmath}
\usepackage{makeidx}
\usepackage{amsfonts}
\usepackage{graphicx,epstopdf}
\usepackage[ansinew]{inputenc}
\usepackage[usenames,dvipsnames]{pstricks}
\usepackage{subfigure}
\usepackage{pdfpages}
\usepackage{epsfig}
\usepackage{pst-grad}
\usepackage{pst-plot}
\usepackage[colorlinks,hyperindex]{hyperref}

\hypersetup{
colorlinks,
citecolor=blue,
linkcolor=black,
urlcolor=black,
}

\newcommand{\be}{\begin{equation}}
\newcommand{\ee}{\end{equation}}

\newcommand{\beq}{\begin{eqnarray}}
\newcommand{\eeq}{\end{eqnarray}}

\begin{document}

\title{Cosmological scenarios from multiquintessence}
\author{R.A.C. Correa}
\email{fis04132@gmail.com}
\affiliation{SISSA-Scuola Internazionale Superiore di Studi Avanzati, via Bonomea, 265, I-34136 Trieste, Italy}
\affiliation{ITA - Instituto Tecnol\'ogico de Aeron\'autica - Departamento de F\'isica,
12228-900, S\~ao Jos\'e dos Campos, S\~ao Paulo, Brazil}
\author{P.H.R.S. Moraes}
\email{moraes.phrs@gmail.com}
\affiliation{ITA - Instituto Tecnol\'ogico de Aeron\'autica - Departamento de F\'isica,
12228-900, S\~ao Jos\'e dos Campos, S\~ao Paulo, Brazil}
\author{A. de Souza Dutra}
\email{dutra@feg.unesp.br}
\affiliation{UNESP, Universidade Estadual Paulista, 12516-410, Guaratinguet\'a, SP, Brazil}
\author{J.R.L. Santos}
\email{joaorafael@df.ufcg.edu.br}
\affiliation{UFCG -  Universidade Federal de Campina Grande - Unidade Acad\^emica de F\'isica, Caixa Postal 10071,
58109-970,  Campina Grande, Para\'iba, Brazil}
\author{W. de Paula}
\email{lspwayne@gmail.com}
\affiliation{Instituto Tecnol\'ogico de Aeron\'autica-ITA- Departamento de F\'isica,
12228-900, S\~ao Jos\'e dos Campos, S\~ao Paulo, Brazil}

\begin{abstract}
In this work we derive and analyse cosmological scenarios coming from
multi-component scalar field models. We consider a direct sum of a sine-Gordon with a Z2 model, and also a combination of those with a BNRT model. \textcolor{black}{Moreover, we work with a modified version of the BNRT model, which breaks the Z2 x Z2 symmetry of the original BNRT potential, coupled with the sine-Gordon and with the standard Z2 models.} We show that our approach can be straightforwardly elevated to $N$ fields. All the computations are made analytically and some
parameters restriction is put forward in order to get in touch with complete %
and realistic cosmological scenarios. 
\end{abstract}

\maketitle

\section{Introduction}

In standard cosmology model (SM), the quantum vacuum energy is the
responsible for the cosmic acceleration observationally predicted \cite%
{riess/1998,perlmutter/1999}. It enters the field equations of General
Relativity in the form of a cosmological constant \textcolor{black}{(CC)}.
However, there is a huge discrepancy between the theoretically predicted
value for the vacuum quantum energy obtained via Particle Physics \cite%
{weinberg/1989} and via observations \cite{hinshaw/2013}. Such
a discrepancy yields the consideration of alternative gravity theories, from
which healthy cosmological models can be derived. Among the alternatives we
quote $f(R)$ and $f(R,T)$ theories \cite{de_felice/2010,nojiri/2006,harko/2011,ms/2017,ms/2016}.

Another possibility comes from scalar field models, named quintessence, in
which it is assumed that the Universe dynamics is governed by some scalar
field $\phi$. Depending on the form of the scalar field time evolution and
on its potential, it is possible to describe the accelerated phases of the
Universe dynamics, namely dark energy and inflationary eras \cite%
{faraoni/2000,elizalde/2008}\textcolor{black}{, without the need of a CC}.
Quintessence models with two scalar fields driving the Universe dynamics
have also generated well-behaved cosmological models \cite%
{khurshudyan/2014,bento/2002,PrdJoaoPedro}.

In the present work we derive cosmological scenarios from N-field
quintessence models, which we call multi-quintessence models. In order to
obtain the scalar field solutions we apply the first order formalism \cite%
{PrdJoaoPedro,bglm,adref}. It is important to remark that, despite the superpotential method was proposed some decades ago, it has been very efficient in solving a large number of physical problems. For instance, we can observe the application of such a method in systems of coupled scalar fields \cite{bazeia}, in
supersymmetric theories \cite{ad4}, in systems of two real scalar fields 
\cite{bnrt}, in Lorentz and CPT breaking
systems \cite{r3}, and in fermionic bound states in distinct kinklike
backgrounds \cite{bazeia12}. 

Here, the solutions for the cosmological parameters
will be compared with observations and we will show that for some values of the free
parameters of the models, it is possible to draw a complete cosmological
scenario, able to describe all the different stages of the
universe evolution.

The fundamental motivation for choosing potentials with more than one scalar field, like those which will be studied in the present work, comes from the fact that using such models, a better observational fitting can be found. It is worth to highlight that in some recent papers \cite{JCAP-2014-044, PRL-2014-161302, PRL-2015-031301}, it was shown that inflationary models based on more than one scalar field are highly compatible with the recent Planck satellite constraints on the scalar index, $n_s$, and the tensor-to-scalar ratio for cosmological density perturbations, $r_T$. In addition, theories with multiple fields open a new window to obtain new constraints for the cosmological parameters, which cannot be derived from single field models, as pointed very recently in \cite{JCAP-2017-049, JCAP-2018-037}.

The paper is organized as follows: in Section \ref{sec:fof} we describe the
first order formalism. In Section \ref{sec:sfs} we obtain our scalar field
solutions and in Section \ref{sec:cpi} we construct the cosmological
parameters of the models. In Section \ref{sec:cr} we revisit our main
results and present a deep discussion about the physical properties of them.

\section{A brief review of the first-order formalism}

\label{sec:fof}

In this section we show a brief overview about the first-order formalism for
the coupling of scalar fields with gravity \cite{PrdJoaoPedro,bglm}. Let us
begin by writing the action in four-dimensional gravity which is coupled to
a set of real scalar fields $\phi _{i}$~,$~i=1,2,...N$, in the following
form 
\begin{equation}
S\!=\!\int d^{4}x\sqrt{|g|\,}\left[ -\frac{1}{4}R+\frac{1}{2}g_{ab}\nabla
^{a}\phi _{i}\nabla ^{b}\phi _{i}-V(\phi _{i})\right] .  \label{eq.1}
\end{equation}%
We are using $c=4\pi G\!=\!1$ and $g\!=\!det(g_{ab})$, with field, space and
time variables being dimensionless and $R$ stands for the Ricci scalar.
Furthermore, we are adopting the signature of the metric as $(+---)$.
Moreover, as usual, $V(\phi _{i})\equiv V\left( \phi _{1},\phi _{2,}...,\phi
_{N}\right) $ is the potential that describes the theory in terms of a
finite arbitrary number of scalar fields.

We will study the cosmological case, where the metric is the
Friedmann-Robertson-Walker one, responsible for a homogeneous and isotropic
expanding universe, 
\begin{equation}
ds^{2}\!=\!dt^{2}-a(t)^{2}\left[ \frac{dr^{2}}{1-kr^{2}}+r^{2}\left( d\theta
^{2}+\sin\theta^{2}d\varphi ^{2}\right) \right] ,  \label{eq3}
\end{equation}%
with $k$ being the curvature parameter, which will be taken as zero, in
accord with observations \cite{hinshaw/2013}. From this metric, the
equations relating the Hubble and the expanding factors are%
\begin{equation}
H^{2}=\frac{2}{3}\rho -\frac{k}{a^{2}};~H\equiv \frac{\dot{a}}{a}
\end{equation}

\noindent and%
\begin{equation}
\frac{\ddot{a}}{a}=-\frac{1}{3}\left( \rho +3~p\right) ;~\bar{q}=\frac{\ddot{%
a}~a}{\dot{a}^{2}}=1+\frac{\dot{H}}{H^{2}},
\end{equation}

\noindent with dots representing time derivatives and $\rho $ and $p$ being
the energy density and pressure of the system, respectively. Furthermore, it
is very useful to work with the equation of state (EoS) parameter 
\begin{equation}
\omega =\frac{p}{\rho }.
\end{equation}

Let us assume that the fields $\phi _{i}$ also depend only on $t$, that is $%
\phi _{i}=\phi _{i}(t)$. Thus, from action (\ref{eq.1}) we can obtain the
corresponding equations of motion for the scalar fields 
\begin{equation}
\ddot{\phi}_{i}+3H\!~\dot{\phi}_{i}+\!V_{\phi _{i}}=0,  \label{eq.4}
\end{equation}%
with $V_{\phi _{i}}\!\equiv \!dV/d\phi _{i}$.

Considering that the potential is given by%
\begin{equation}
V\left( \phi _{i}\right) =\frac{3}{2}W\left( \phi _{1},\phi _{2,}...,\phi
_{N}\right) ^{2}-\frac{1}{2}\sum_{i=1}^{N}W_{\phi _{i}}^{2},
\end{equation}%
with $W(\phi_i)$ being a superpotential and $W_{\phi _{i}}\equiv dW/d\phi_i$%
, one can perceive that even if we use a superpotential which is a direct
sum of one scalar field functions, the resulting field potential is
non-trivial as it could be seen in braneworld scenarios \cite%
{gustavodutrahoff}, for example. The energy density and pressure in this formalism are
written as 
\begin{eqnarray}
\rho &=&\sum_{j=1}^{N}\frac{\dot{\phi}_{j}^{2}}{3}+~\frac{3}{2}W\left( \phi
_{1},\phi _{2,}...,\phi _{N}\right) ^{2}-\frac{1}{2}\sum_{i=1}^{N}W_{\phi
_{i}}^{2},~ \\
p &=&\sum_{j=1}^{N}\frac{\dot{\phi}_{j}^{2}}{3}-~\frac{3}{2}W\left( \phi
_{1},\phi _{2,}...,\phi _{N}\right) ^{2}+\frac{1}{2}\sum_{i=1}^{N}W_{\phi
_{i}}^{2}.
\end{eqnarray}

Furthermore, in terms of the superpotential, the differential equations of
motion can be cast in the first order form when we use the identification $%
H=-~W\left( \phi _{1},\phi _{2,}...,\phi _{N}\right) $, and are written as%
\begin{equation}
\dot{\phi}_{i}=W_{\phi _{i}},~i=1,...,N.  \label{q1}
\end{equation}

From this approach, it was shown in Ref.\cite{gustavodutrahoff} that some
braneworld scenarios allow to find interesting features, such as split brane
mechanism and asymmetric warp factor shape, which is essential to address
the hierarchy problem in the context of thick brane scenarios. As
demonstrated by the authors, the method is not restricted to the combination
of one-field nonlinear superpotentials. In fact one could combine two or
more superpotentials where two or more coupled scalar fields appear non
trivially.

In the next section, we will apply the approach described above to study
cosmological scenarios coming from multi-component scalar field models.

Our main goal in the next sections will be to generate two periods of accelerated expansion - one
due to the inflation and the second due to dark energy. It is well known
that the cosmological constant can be used for generating such scenarios. However, since the inflation era must end, the
cosmological constant cannot be used for this purpose, beyond the fact that
it needs a great fine tuning. On this regard, the so called quintessence
approach \cite{PrdJoaoPedro, bglm} is an alternative way to obtain an
accelerated expansion era and it is associated to the movement of the
field(s) from one vacuum of the theory to another. Usually, exact models
generate just one of these accelerated eras \cite{PrdJoaoPedro, bglm}. Here
we are going to show that it is possible to get qualitatively these two
epochs of acceleration and\textcolor{black}{, remarkably, also} the
intermediate decelerated eras from the combination of two or more models. %
\textcolor{black}{It is worth remarking that the accelerated phases will be
obtained with no need of a CC, in such a way that our models evade the
cosmic coincidence and cosmological constant problems
\cite{zlatev/1999,arkani-hamed/2000,peebles/2003,padmanabhan/2003}.}

\section{Scalar field solutions}

\label{sec:sfs}

Now we go further in the approach, by presenting some 
analytical examples.

\subsection{Model I}

We will start by considering a superpotential which is a direct sum of a
sine-Gordon \cite{rajaraman} and a $Z_{2}$ model \cite{vaca}, as:%
\begin{equation}
W\left( \phi _{1},\phi _{2}\right) \equiv \lambda _{1}\left( \phi _{1}-\frac{%
\phi _{1}^{3}}{3}\right) +\lambda _{2}~\sin \phi _{2}+\alpha _{1}.
\label{q2}
\end{equation}

\noindent where $\lambda_1$, $\lambda_2$ and $\alpha _{1}$ are arbitrary constant.

It is worth to emphasize that the $Z_{2}$ model is the most fundamental
example where we can find the so-called $Z_{2}$ kink. In this context, the
Lagrangian is invariant under the transformation $\phi \rightarrow -\phi $
and hence there is a reflectional $Z_{2}$ symmetry. On the other hand, the
sine-Gordon model is invariant under $\phi \rightarrow \phi +2\pi n$, where $%
n$ is any integer, and thus possesses $Z$ symmetry. 

The motivation to study this first example, where we have a mixture of sine-Gordon and $Z_{2}$
models, comes from the fact that those are found in many areas of Physics,
including condensed matter physics, field theory, cosmology, among many
others. 

For instance, as it is well known, whenever we have a potential with
two or more degenerate minima, one can find different vacua at different
portions of the space. Thus, one can find domain walls connecting such
regions. In this framework, since the Universe is expanding and cooling,
cosmic phase transitions can occur and domain walls can form \cite{vaca}.

Using Eq.(\ref{q1}), we obtain the following differential equations%
\begin{eqnarray}
\dot{\phi}_{1} &=&\lambda _{1}\left( 1-\phi _{1}^{2}\right) ,  \label{q3} \\
\dot{\phi}_{2} &=&\lambda _{2}\cos\phi_{2}.  \label{q4}
\end{eqnarray}
The solutions of Eqs.(\ref{q3}) and (\ref{q4}) are, respectively, given by 
\begin{eqnarray}
\phi _{1}(t) &=&\tanh \left( \lambda _{1}t\right) ,  \label{q5} \\
\phi _{2}(t) &=&\arctan \left[ \sinh \left( \lambda _{2}t\right) \right] .
\label{q6}
\end{eqnarray}

\subsection{Model II}

A second example can be represented as a combination of $Z_{2}$, sine-Gordon
and BNRT models \cite{bnrt}. In this case, the superpotential
can be written in the following form 
\begin{eqnarray}
&&\left. W\left( \phi _{1},\phi _{2},\phi _{3},\phi _{4}\right) =\lambda
_{1}\left( \phi _{1}-\frac{\phi _{1}^{3}}{3}\right) +\lambda _{2}~\sin \phi
_{2}\right.  \notag \\
&&\left. -\lambda _{3}\phi _{3}+\frac{\lambda _{3}}{3}\phi _{3}^{3}+\mu
_{3}\phi _{3}\phi _{4}^{2}+\alpha _{2}.\right.  \label{q7}
\end{eqnarray}

\noindent where $\lambda_3,\mu_3$ and $\alpha _{2}$ is an arbitrary constant.

Let us remark that the BNRT model \cite{bnrt} is used for modeling a great
number of systems \cite{ad1,ad2,ad3,ad4,ad5}. This model admits a variety of
kink-like solutions and it has been shown to give rise to degenerate and
critical branes \cite{r1}, traveling solitons \cite{r2, r3} and BPS and
non-BPS defects \cite{r5}. Moreover, using the so-called configurational
entropy \cite{r7}, it was shown in Ref.\cite{r9}\ that the model has a rich
structure, which allows the generation of double-kink configurations.

From Eq.(\ref{q7}), we have the following first-order equations%
\begin{eqnarray}
\dot{\phi}_{1} &=&\lambda _{1}\left( 1-\phi _{1}^{2}\right) ,  \label{q8} \\
\dot{\phi}_{2} &=&\lambda _{2}\cos\phi_{2},  \label{q9} \\
\dot{\phi}_{3} &=&-\lambda _{3}\left( 1-\phi _{3}^{2}\right) +\mu _{3}\phi
_{4}^{2},  \label{q10} \\
\dot{\phi}_{4} &=&2\mu _{3}\phi _{3}\phi _{4}.  \label{q11}
\end{eqnarray}

We can see that Eqs.(\ref{q8}) and (\ref{q9}) also have Eqs.(\ref{q5}) and (%
\ref{q6}) as solutions, respectively. On the other hand, Eqs.(\ref{q10}) and
(\ref{q11}) are coupled. In this case, it has been found in Ref.\cite{dutra1}
a class of analytical solutions for this system, which was named Degenerate
Bloch Walls. It was shown that there are two situations with exact
classical solutions. The first one is given by

\begin{eqnarray}
\phi _{3}^{(1)}(t) &=&\frac{\left( \sqrt{c_{0}^{2}-4}\right) \sinh (2\mu
_{3}t)}{\left( \sqrt{c_{0}^{2}-4}\right) \cosh (2\mu _{3}t)-c_{0}},
\label{q12} \\
\phi _{4}^{(1)}(t) &=&\frac{2}{\left( \sqrt{c_{0}^{2}-4}\right) \cosh (2\mu
_{3}t)-c_{0}},  \label{q13}
\end{eqnarray}

\noindent where $c_{0}<-2$ and $\lambda _{3}=\mu _{3}$.

For the second class of solutions, $c_{0}<1/16$ and $\lambda _{3}=4\mu _{3}$%
. In this case, we have 
\begin{eqnarray}
\phi _{3}^{(2)}(t) &=&\frac{\left( \sqrt{1-16c_{0}}\right) \sinh (4\mu _{3}t)%
}{\left( \sqrt{1-16c_{0}}\right) \cosh (4\mu _{3}t)+1},  \label{sol1} \\
\phi _{4}^{(2)}(t) &=&-\frac{2}{\left( \sqrt{1-16c_{0}}\right) \cosh (4\mu
_{3}t)+1}.  \label{sol2}
\end{eqnarray}

\begin{figure}[h]
\includegraphics[width=0.9\columnwidth]{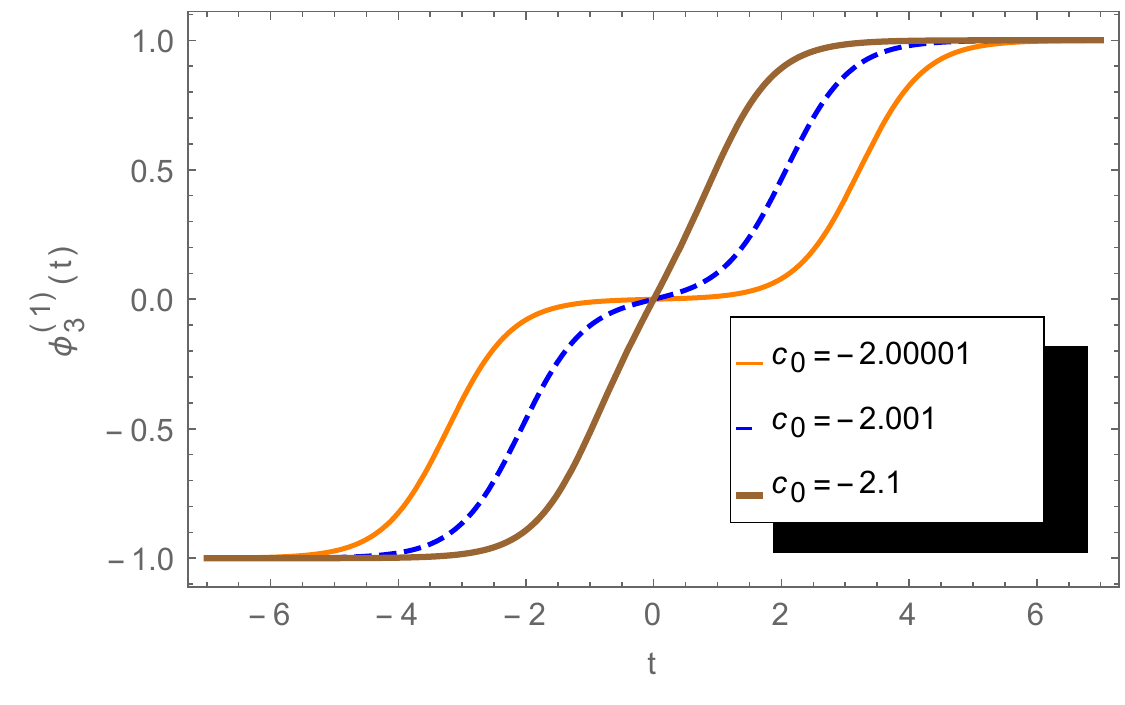}
\caption{Profiles of the DBW solutions $\protect\phi _{3}^{(1)}(t)$ 
\textcolor{black}{(Eq.(\ref{q12}))} with $\protect\mu=1$.}
\label{FIG1}
\end{figure}

\begin{figure}[h]
\includegraphics[width=0.9\columnwidth]{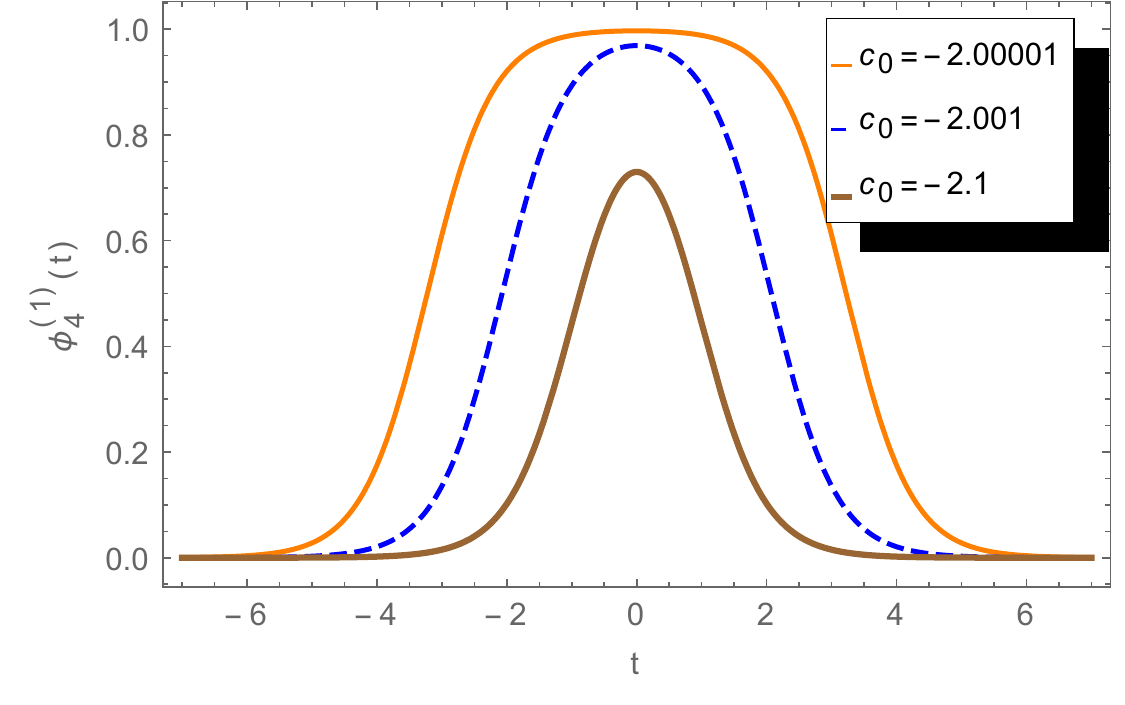}
\caption{Profiles of the DBW solutions $\protect\phi _{4}^{(1)}(t)$ 
\textcolor{black}{(Eq.(\ref{q13}))} with $\protect\mu=1$.}
\label{FIG2}
\end{figure}

In Figs. \ref{FIG1} and \ref{FIG2} below we show some typical profiles of the degenerate
bloch walls solutions. Note that the two-kink solution $\phi _{3}^{(1)}$ arises only for values of $%
c_{0}$ close to the critical value, $c_{0}^{crit}=-2$. For the same values
of $c_{0}$, the corresponding lump-like solutions for $\phi _{4}^{(1)}$
exhibit a flat top, which disappears as one moves away from $c_{0}^{crit}$.
A remarkable feature of these solutions is that their BPS energies \cite%
{rajaraman} are degenerate with respect to $c_{0}$, i.e., for any value of $%
c_{0}$, the energy is given by $E_{BPS}=4\lambda /3$.

\subsection{Model III}

\textcolor{black}{As a third example, we consider the following superpotential}
\begin{eqnarray}
\label{q13_1}
 W &=&\lambda_1\,\left(\phi_1-\frac{\phi_1^{\,3}}{3}\right)+\lambda_2\,\sin\,\phi_2 \\ \nonumber
 &&
 -\lambda_3\,\frac{\phi_3^{\,3}}{3}-\phi_3^{\,2}\,\phi_4+\phi_4-\frac{\phi_4^{\,3}}{3}+\alpha_3\,,
\end{eqnarray}
\textcolor{black}{with $\alpha_3$ being constant. Here, the $\phi_3$-$\phi_4$ dependence was inspired in a model studied by \cite{britodutra}, where the authors found interesting asymmetric two-kink analytical solutions. }

\textcolor{black}{We can directly determine that the first-order differential equations for fields $\phi_3$ and $\phi_4$ are}
\begin{eqnarray}
&&
\dot{\phi}_3=-\lambda_3\,\phi_3 ^{\,2}-2\, \phi_4\, \phi_3\,, \label{q14}  \\
&&
\dot{\phi}_4=-\phi_4 ^{\,2}-\phi_3^{\,2} +1\, \label{q15}.
\end{eqnarray}

\textcolor{black}{Moreover, as we saw in the second example, $\phi_1$ and $\phi_2$ obey the first-order equations $(\ref{q8})$ and $(\ref{q9})$, respectively. The analytical solutions which satisfy equations $(\ref{q14})$ and $(\ref{q15})$ have the forms}

\begin{figure}[h]
\includegraphics[width=0.9\columnwidth]{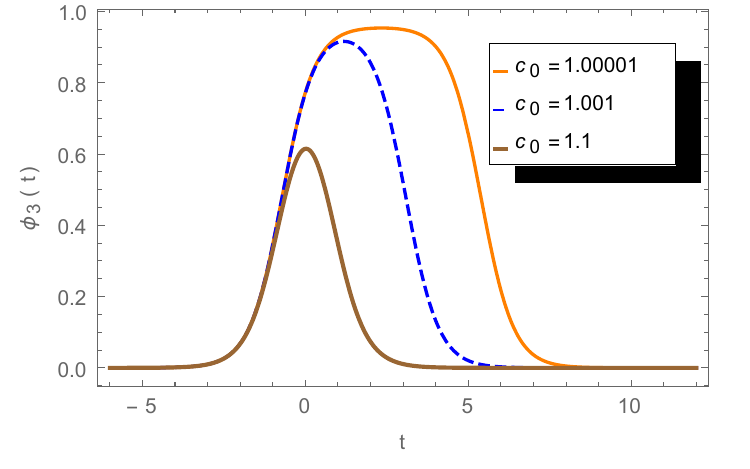}
\caption{Profiles of solutions $\protect\phi _{3}(t)$ 
\textcolor{black}{(Eq.(\ref{q16}))} with $\protect\lambda_3=0.6$.}
\label{FIG3}
\end{figure}

\begin{figure}[h]
\includegraphics[width=0.9\columnwidth]{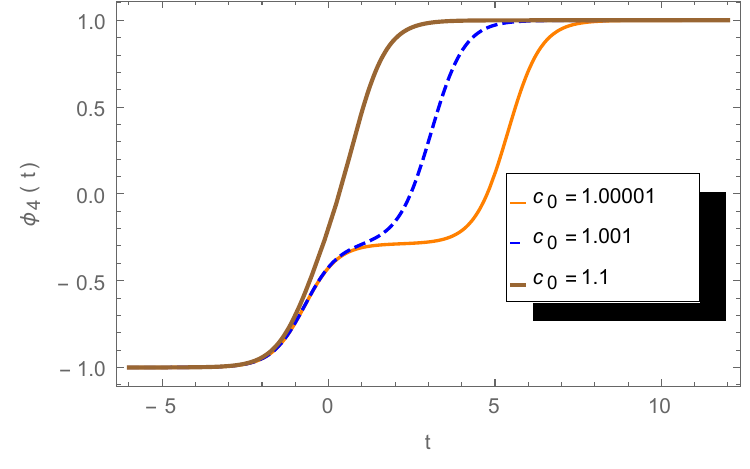}
\caption{Profiles of solutions $\protect\phi _{4}(t)$ 
\textcolor{black}{(Eq.(\ref{q17}))} with $\protect\lambda_3=0.6$.}
\label{FIG4}
\end{figure}

\begin{widetext}

\begin{equation}
\label{q16}
 \phi_3=4\,\left\{\left(c_0^2 \left(\lambda_3^2+4\right)-\lambda_3^2-5\right) \sinh (2\,t)+\left(c_0^2 \left(\lambda_3^2+4\right)-\lambda_3^2-3\right) \cosh (2\,t)+2 c_0 \sqrt{\lambda_3^2+4}\right\}^{\,-1}\,,
\end{equation}
\begin{eqnarray}
\label{q17}
\phi_4 & = & \left\{\left(c_0^2 \left(\lambda_3^2+4\right)-\lambda_3^2-3\right) \sinh (2\,t)+\left(c_0^2 \left(\lambda_3^2+4\right)-\lambda_3^2-5\right) \cosh (2\,t)-2 \lambda_3\right\}\\ \nonumber
&&
\left\{\left(c_0^2 \left(\lambda_3^2+4\right)-\lambda_3^2-5\right) \sinh (2\,t)+\left(c_0^2 \left(\lambda_3^2+4\right)-\lambda_3^2-3\right) \cosh (2\,t)+2 c_0 \sqrt{\lambda_3^2+4}\right\}^{\,-1}\,,
\end{eqnarray}

\end{widetext}
\noindent \textcolor{black}{whose profiles are show in Figs. \ref{FIG3} and \ref{FIG4}, respectively. It is relevant to point that Fig.  \ref{FIG4} unveils the asymmetric two-kink profile when we consider values close to $c_0^{crit}=1$.}

\section{Cosmological parameters solutions}

\label{sec:cpi}

Below we plot the cosmological solutions for $H(t)$, $\omega(t)$ and $\bar{q}%
(t)$, defined in Section \ref{sec:fof}, for the three models described above, with superpotential (\ref{q2})
representing what we are calling \textit{Model I}, while superpotential (\ref{q7}%
) represents \textit{Model II} and finally superpotential $(\ref{q13_1})$ stands for \textit{Model III}.

\begin{figure}[h]
\includegraphics[width=0.9\columnwidth]{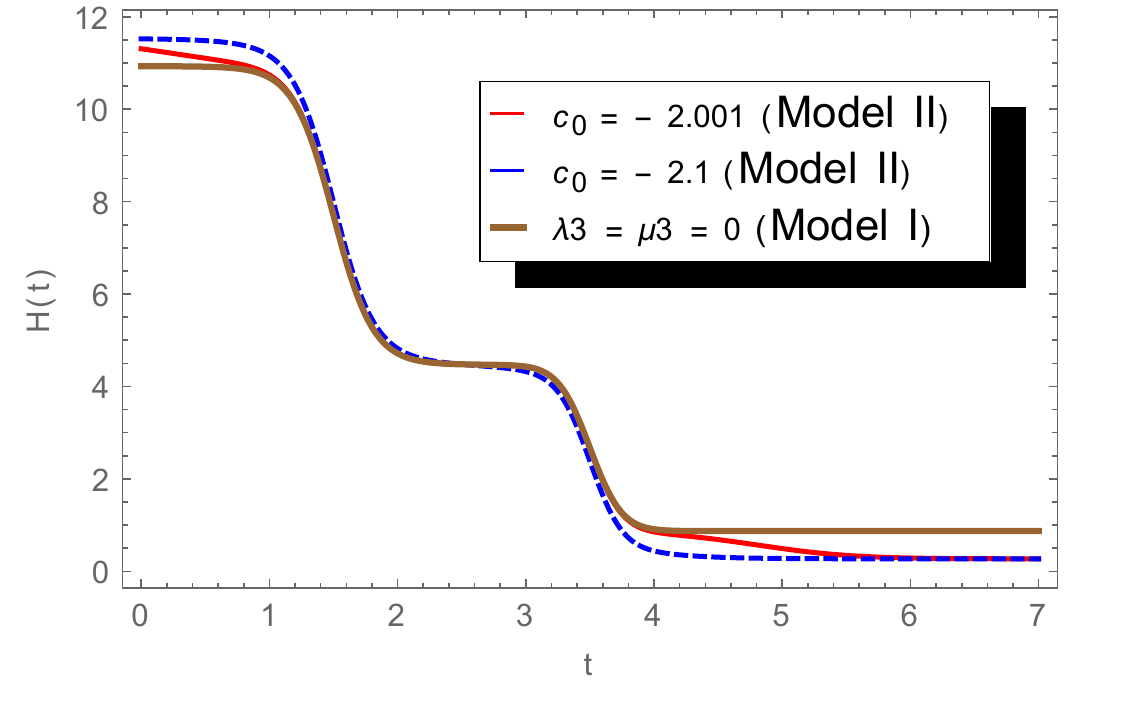}
\caption{Hubble factor for different values of the $c_{0}$ parameter with $%
\protect\lambda _{1}=2.70$, $\protect\lambda _{2}=3.23$, and $\protect\alpha %
_{1}=\alpha_2=-5.9$. The (blue) dotted and (red) thin lines stands for $\protect%
\lambda _{3}=\protect\mu _{3}=0.9$.}
\label{FIG5}
\end{figure}

\begin{figure}[h]
\includegraphics[width=0.9\columnwidth]{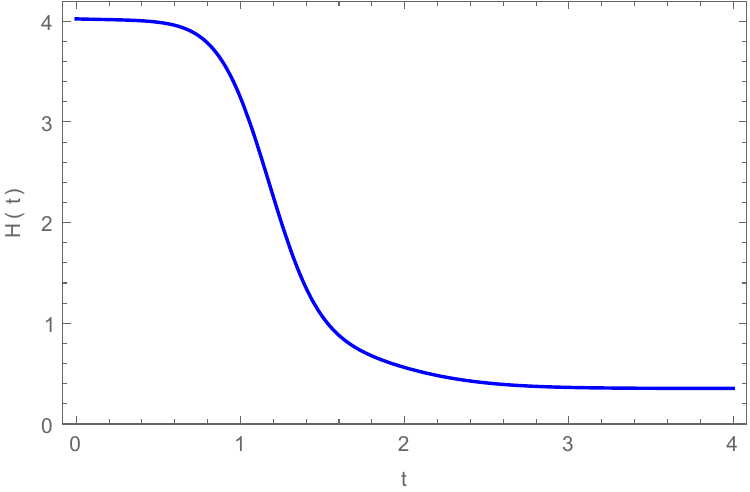}
\caption{Hubble factor for {\it Model III}, with $c_{0}=1.00015$, $%
\protect\lambda _{1}=2.25$, $\protect\lambda _{2}=4.5$, $\protect\lambda_3=0.00005$, and $\alpha_3=-7.02$}
\label{FIG5_1}
\end{figure}

\begin{figure}[h]
\includegraphics[width=0.9\columnwidth]{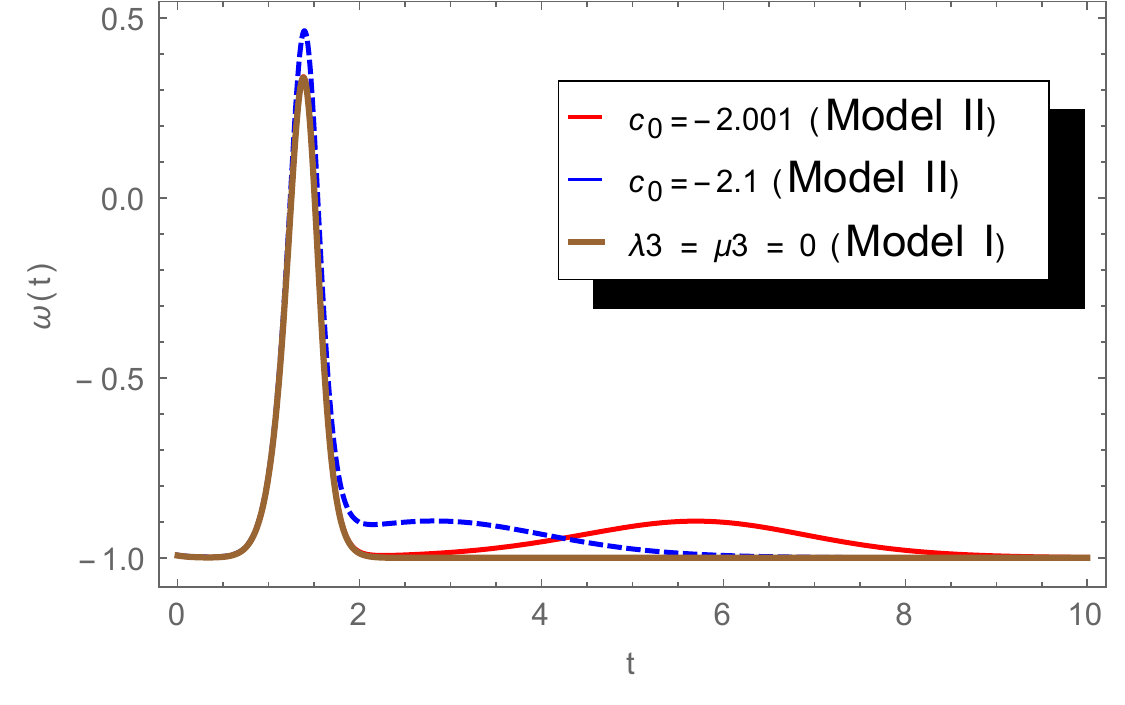}
\caption{Behavior of the equation of state with $\protect\lambda _{1}=2.70$, 
$\protect\lambda _{2}=3.23$, and $\protect\alpha _{1}=\alpha_2=-5.9$. The (blue)
dotted and (red) thin lines stands for $\protect\lambda _{3}=\protect\mu %
_{3}=0.9$.}
\label{FIG6}
\end{figure}

\begin{figure}[h]
\includegraphics[width=0.9\columnwidth]{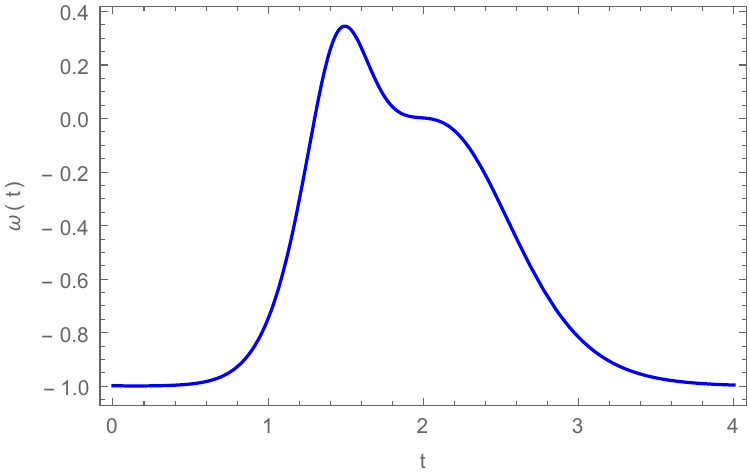}
\caption{Behavior of the equation of state for {\it Model III}, with $c_{0}=1.00015$, $%
\protect\lambda _{1}=2.25$, $\protect\lambda _{2}=4.5$, $\protect\lambda_3=0.00005$, and $\alpha_3=-7.02$}
\label{FIG6_1}
\end{figure}

\begin{figure}[h]
\includegraphics[width=0.9\columnwidth]{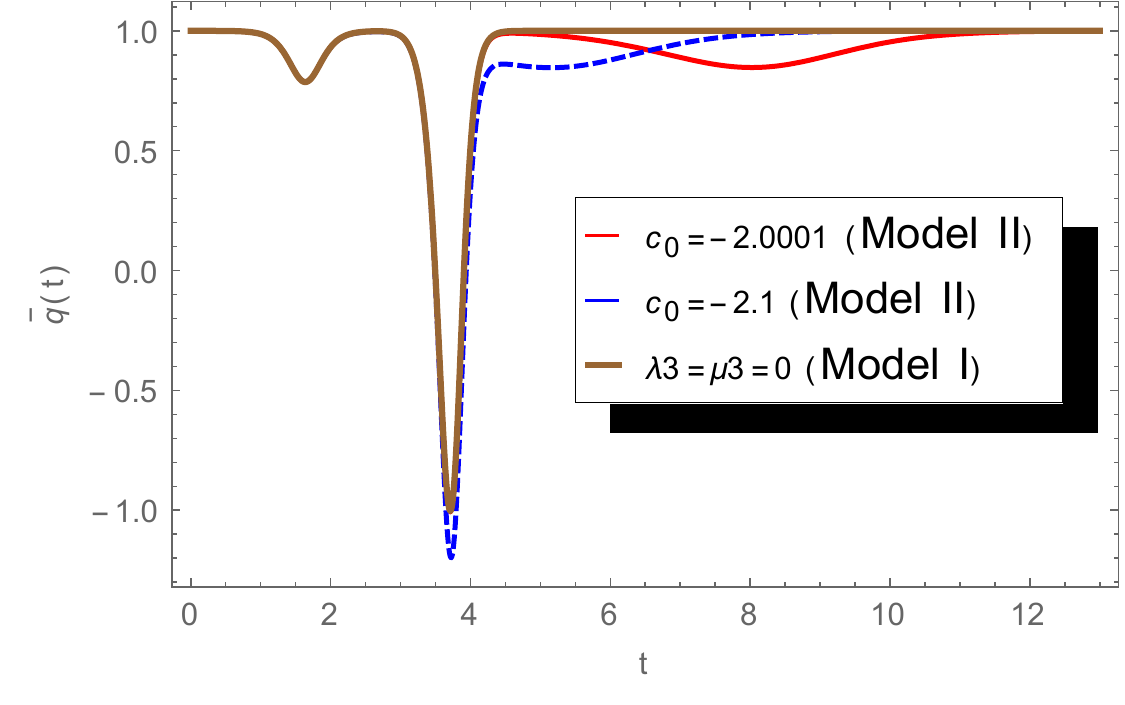}
\caption{Acceleration parameter evolution for different values of the $c_{0}$
parameter with $\protect\lambda _{1}=2.70$, $\protect\lambda _{2}=3.23$, and 
$\protect\alpha _{1}=\alpha_2=-5.9$. The (blue) dotted and (red) thin lines stands
for $\protect\lambda _{3}=\protect\mu _{3}=0.9$.}
\label{FIG7}
\end{figure}

\begin{figure}[h]
\includegraphics[width=0.9\columnwidth]{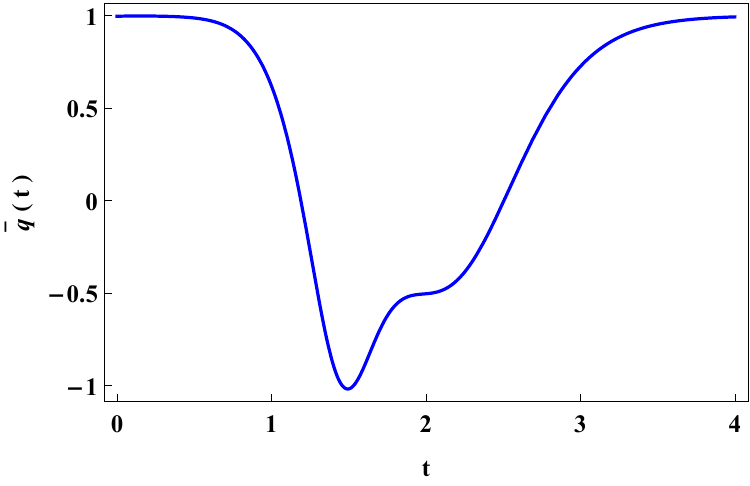}
\caption{Acceleration parameter evolution  for {\it Model III}, with $c_{0}=1.00015$, $%
\protect\lambda _{1}=2.25$, $\protect\lambda _{2}=4.5$, $\protect\lambda_3=0.00005$ and $\alpha_3=-7.02$.}
\label{FIG7_1}
\end{figure}

In the next section, we will interpret Figs.\ref{FIG5}-\ref{FIG7_1} and point the features that
corroborate to the fact that for some chosen values of the free parameters,
we are able to describe a complete cosmological scenario, i.e., one that
includes the inflationary, radiation, matter and dark energy dynamical eras 
\cite{ryden/2003,weinberg/2008}, as well as the transitions among these
stages, in a continuous \textcolor{black}{and analytical} form.

\section{Discussion}

\label{sec:cr}

The main cosmological puzzle nowadays is to predict in a theoretical level
the observed cosmic acceleration. Whether this phenomenon comes from extra
degrees of freedom in the geometrical description \cite{nojiri/2008,hu/2007}
or material content \cite%
{harko/2011,ms/2016,elizalde/2004,amendola/2000,das/2006} of the Universe is
still an open question.

In SM, the cosmic acceleration is considered as due to the presence of the
CC in the Einstein's field equations. As quoted in the Introduction, such a
consideration yields the CC problem. \textcolor{black}{Here, instead, as we
are going to argue below, we were able to predict some features of an
accelerated universe, i.e., $H\sim cte$, $\omega<-1/3$ and $\bar{q}>0$, with
no need of the CC, evading, in this way, the CC problem.} In fact, there is
a number of shortcomings or inconsistencies surrounding SM, as one can also check
in \cite{wilczek/2004,capozziello/2008}, for instance.

Another shortcoming surrounding the SM, barely explored in the literature,
is that within its formalism it is not possible to describe the whole
Universe dynamics \textcolor{black}{in a both analytical and continuous form,
as we discuss below}.

We know that since the Big-Bang the Universe is expanding, however, the
dynamical features of this expansion are not the same since then. Right
after the Big-Bang, the Universe passed through an accelerated regime of
expansion, named inflation (or inflationary era) \cite{guth/1981}. The
inflaton, which is named the field responsible for bringing such an abrupt
dynamics to the Universe decays rapidly in radiation. After radiation, (dark
and barionic) matter and dark energy, subsequently, dominate the Universe
dynamics \cite{ryden/2003}. To say that SM is unable to analytically
describe the dynamics of the Universe in a continuous form means that the
cosmological parameters of SM, such as the EoS and acceleration parameters,
are analysed separately for each of those stages.

In other words, in SM, one is unable to find an analytical solution, say,
for the EoS parameter, which is able to describe the inflationary,
radiation, matter and dark energy eras, and the transitions among them, in a
continuous and self-consistent form. Rather, when solving the standard
Friedmann equations, it is admissible to know the EoS of the dominant
component of the Universe dynamics at a specific time, be the inflaton,
radiation, matter or dark energy.

Let us clarify this statement by writing the (standard) Friedmann equations
and the energy-momentum tensor conservation equation of SM (for a flat
space-time):

\begin{eqnarray}
\left(\frac{\dot{a}}{a}\right)^{2}=\frac{2}{3}\rho,  \label{d1} \\
\frac{\ddot{a}}{a}+\frac{1}{2}\left(\frac{\dot{a}}{a}\right)^{2}=-\omega\rho,
\label{d2} \\
\dot{\rho}+3\frac{\dot{a}}{a}(1+\omega)\rho=0,  \label{d3}
\end{eqnarray}
in which it was already assumed the EoS $p=\omega\rho$. We can see from (\ref%
{d1})-(\ref{d3}) a set of three equations with three unknowns: $a$, $\rho$
and $\omega$. However, Eq.(\ref{d3}) can also be obtained by rearranging (%
\ref{d1}) and (\ref{d2}). In this way, two of the three equations are
independent, what makes the system insoluble.

What one makes, then, is to assume a specific value for the EoS $\omega$ and
solve the Friedmann equations for $a$ and $\rho$. Usually, one takes $%
\omega=1/3$ and $\omega=0$ for radiation and matter components,
respectively. On the other hand, the accelerated stages, namely inflationary
and dark energy eras, can be described by an exotic fluid of negative EoS,
specifically with $\omega<-1/3$ \cite{ryden/2003}.

Anyhow, as argued above, so far there is no analytical approach in SM able to describe a
unique functional form for $\omega(t)$ that describes the different stages
of the Universe as well as the respective transitions.

In the present article, from the application of the first order formalism in
Section \ref{sec:sfs}, we were able to construct the cosmological scenarios
of Section \ref{sec:cpi}. In the next paragraphs we will check the
reliability of these scenarios and its relation with the values of the free
parameters of the models. For now, it is important to remark that we
obtained a complete set of exact solutions for the studied models, which
display double and single-kink configurations, as we can see from Figs.\ref{FIG1}-\ref{FIG4}. In fact, such models are very important in applications that include
Bloch branes, Skyrmions, Yang-Mills, Q-balls, oscillons and various
superstring-motivated theories.

Let us start by analysing the Hubble factor (Figs.\ref{FIG5}-\ref{FIG5_1}). Qualitatively, all
curves are in accord with SM in the sense that $H\propto t_H^{-1}$, with $%
t_H $ being the Hubble time. Still in agreement with SM, $H$ is restricted
to positive values ($H<0$ would imply a contracting universe).

One might wonder the reason why $H(t)\rightarrow constant$ for high values
of time. In order to interpret such a behavior of $H(t)$, one should keep in
mind that $H=\dot{a}/a$ and that an accelerated expansion might be described
by an exponential of $t$ as $a\sim e^{H_0t}$, with $H_0$ a constant (in this
case, the present value of $H(t)$). Such a solution for $a(t)$ yields a
constant behavior for $H(t)$, therefore $H=constant$ in Figs.\ref{FIG5}-\ref{FIG5_1} for high
values of time stands for the recent regime of accelerated expansion the
Universe is passing through.

Figs.\ref{FIG6}-\ref{FIG6_1} show the time evolution of the EoS parameter of Models I, II and III. As
described some paragraphs above, in SM one assumes to know the values of
this parameter for different eras of the Universe, but there is no mechanism
capable to predict the functional form of $\omega(t)$ %
\textcolor{black}{through the whole universe evolution}. Here, on the other
hand, we have obtained from Model III a solution to $\omega(t)$ and this is
describing the whole evolution of the Universe dynamics as we will argue in
the following.

Firstly we note in both figures that $\omega$ starts with negative values, specifically,
when $t\sim0$, $\omega\sim-1$, independently of the model. This is in
agreement with some constraints that were recently put to the inflationary
EoS \cite{bamba/2013,ackerman/2011,aldrovani/2008}. Note that this behavior
is reflected in the curves of $H(t)$ for small values of time ($%
t\rightarrow0\Rightarrow H(t)\rightarrow constant$).

After describing inflation, $\omega$ tends to positive values. Note that for
the values chosen in Figs.\ref{FIG6}-\ref{FIG6_1} for the constants of the model, $\omega$ has its
maximum value around $1/3$. Indeed, that is the maximum value the EoS of the
Universe should assume according to SM, as mentioned above. In this way, our
models predict a smooth transition between inflationary and radiation eras,
which is not trivial to be obtained, even with alternative models (check
Refs.\cite{lima/2013,perico/2013} for some content about the so-called
``graceful exit'').

The EoS parameter, then, starts to decrease, eventually re-assuming negative
values. Note that for higher values of time, $\omega\rightarrow-1$, which is
related with the cosmic acceleration and is in accordance with the present
value of the EoS, obtained via observation of fluctuations of the cosmic
microwave background temperature ($\omega=-1.073^{+0.090}_{-0.089}$) \cite%
{hinshaw/2013}.

\textcolor{black}{It is important to remark that the Model III was able to remarkably allow a matter-dominated era between radiation and dark energy-dominated stages. This is represented by a plateau-like behaviour for $\omega$ when it reaches $0$. It can be seen that $\omega$ remains null for a non-negligible period of time. Such a value for $\omega$ describes the matter-dominated era and the period of time in which it remains zero would allow matter (star, galaxies, clusters of galaxies etc.) to be formed in the Universe.}

Let us now check Figs.\ref{FIG7}-\ref{FIG7_1}, in which the acceleration parameter is depicted. For
small values of time, $\bar{q}>0$, which stands for an accelerated stage of
the Universe expansion. According to the argumentations above, we can
conclude that the behavior of $\bar{q}$ for small values of $t$ is nothing
but the description of the inflationary era.

As time passes by, $\bar{q}$ tends to negative values, which describe a
decelerated expansion of the Universe. As also argued above, we know the
dynamics of the Universe expansion changes with time. After inflationary
era, the Universe expanded in a decelerated way, during radiation and matter
epochs. Those are described by the period of time in which $\bar{q}<0$ in
Fig.\ref{FIG7}.

Then, $\bar{q}$ turns positive, passes through its current %
\textcolor{black}{observationally predicted} value \cite{hinshaw/2013},
describing the present cosmic acceleration, and tends to $1$, asymptotically
predicting, in this way, a de Sitter-like universe in the far future (the
same prediction is appreciated in Fig.\ref{FIG5}, since the de Sitter universe is
described by $H\sim cte$).

\textcolor{black}{We also remark that Fig.\ref{FIG7_1} shows a period of matter-dominated dynamics before $\bar{q}$ returns to positive values. This is represented by a plateau behaviour at $\bar{q}=-0.5$. Remarkably, according to SM, this is the value of the acceleration parameter at the matter-dominated era of the Universe evolution \cite{ryden/2003}.}

In this way, our multi-quintessence models have shown to be able to provide
a complete \textcolor{black}{analytical} cosmological scenario%
\textcolor{black}{, i.e., a scenario which describes the inflationary,
radiation, matter and dark energy eras, as well as the transitions among
them, in a continuous form. Such complete and analytical scenarios can
hardly be found in the literature, as one can check the $f(R,T^\phi)$ and
$\Lambda(t)$ models \cite{ms/2016,lima/2013}, with $T^\phi$ being the trace
of the energy-momentum tensor of a scalar field and $\Lambda(t)$  a decaying
cosmological ``constant''}.

Deeper numerical fittings possibly will need additional restrictions of the
potential parameters or a higher number of exact fields in the model. It is
interesting to note that the integration constant appearing in the BNRT
models can be used for tunings without changing the values of the potential
parameters, which make them interesting in the sense that the model becomes
more robust due to the possibility of avoiding fine tunings of the potential
parameters.

An important consequence of our investigations is the fact that the cosmological parameters are not strongly bounded by the free parameters of the model. If we consider small variations in the values chosen for the parameters, we will obtain similar behaviors for the cosmological parameters, characterizing a robust result for the model under analysis. 

On the other hand, by altering the form of the potential, different cosmological scenarios may be expected. In fact, we had previously taken other potentials into account, but those did not lead to the remarkable features presented in the article, that is, a complete scenario for the universe dynamics evolution. However, further choices for the potential may be investigated in the future.

\acknowledgments R.A.C. Correa thanks to S\~ao Paulo Research Foundation
(FAPESP), grant numbers 2016/03276-5 and 2017/26646-5 for financial support. P.H.R.S. Moraes thanks FAPESP,
Grant No. 2015/08476-0, for financial support. ASD, JRLS, and WP thanks CNPq for financial support.

\end{document}